\documentclass[useAMS,usegraphicx,usenatbib,referee]{mn2e}

\title[Non linear particle acceleration at shock waves]
{Non linear particle acceleration at non-relativistic shock waves
in the presence of self-generated turbulence}
\author[E. Amato and P. Blasi]{E. Amato$^{1}$\thanks{E-mail:
amato@arcetri.astro.it} 
and P. Blasi$^{1}$\thanks{E-mail: blasi@arcetri.astro.it}\\
$^{1}$INAF-Osservatorio Astrofisico di Arcetri, 
Largo E. Fermi, 5, 50125, Firenze, Italy}

\begin{document}

\date{Accepted ----. Received -----}


\maketitle

\label{firstpage}

\begin{abstract}
Particle acceleration at astrophysical shocks may be very efficient
if magnetic scattering is self-generated by the same particles. This
nonlinear process adds to the nonlinear modification of the shock due
to the dynamical reaction of the accelerated particles on the shock. 
Building on a previous general solution of the problem of particle
acceleration with arbitrary diffusion coefficients (\cite{amato1}), we 
present here the first semi-analytical calculation of particle 
acceleration with both effects taken into account at the same time: 
charged particles are accelerated in the background of Alfv\'en waves
that they generate due to the streaming instability, and modify the 
dynamics of the plasma in the shock vicinity. 
\end{abstract}


\section{Introduction}
\label{sec:intro}

Soon after the pioneering papers by \cite{krymskii,bo78,bell78a,bell78b},
introducing the test particle theory of particle acceleration
at collisionless shocks, it became clear that the dynamical 
reaction of the accelerated particles on the plasmas involved 
in the shock formation may not be negligible. It is now clear 
that such reaction may in fact make shocks efficient accelerators
and change quite drastically the predictions of the test particle 
theory. The main consequences of the shock modification induced
by the accelerated particles
can be summarized as follows: 1) a precursor, consisting in a gradual 
braking of the upstream fluid, is created; 2) particles with different
momenta {\it feel} different effective compression factors, which 
reflects in the fact that the spectrum of accelerated particles is
no longer a power law, but rather a concave spectrum, as hard as 
$p^{-3.2}$ at high momenta; 3) the shock becomes less efficient in
heating the background plasma, so that the temperature of the downstream
gas is expected to be lower than predicted through the usual 
Rankine-Hugoniot relations at an unmodified shock front (see 
\cite{drury83,be87,je91,maldru2001} for reviews on different 
aspects of the subject). The reaction of the accelerated particles
has been calculated within different approaches: the so-called 
two-fluid models (\cite{dr_v80,dr_v81}), kinetic models 
(\cite{malkov1,malkov2,blasi1,blasi2}) and numerical
approaches, both Monte Carlo and other simulation procedures 
(\cite{je91,bell87,elli90,ebj95,ebj96,kj97,kj05,jones02}).
In most of these calculations, the diffusion properties of 
the plasma upstream and downstream are provided as an input to the 
problem. This also results in fixing the value of the maximum 
momentum of the accelerated particles. However, one of the well
known and most disturbing problems associated with the mechanism
of particle acceleration at shock fronts is that a substantial 
amount of magnetic scattering of the particles is required
(e.g. \cite{lc83a,lc83b}). In the absence of it, the maximum energy of the
accelerated particles is exceedingly low and uninteresting for 
astrophysical applications (e.g. \cite{blasirev}). \cite{bell78a}
proposed that the streaming instability of cosmic rays could be
responsible for the generation of perturbations in the magnetic
field of an amplitude necessary to provide pitch angle scattering
(and therefore spatial diffusion) of the accelerated particles. 
\cite{lc83b} used this argument to estimate the maximum energy of
particles accelerated at shocks in supernova remnants. In all previous 
works either the shock was considered unmodified, or the diffusion 
coefficients were fixed {\it a priori}, because a comprehensive
theory of particle acceleration was missing. Recently, \cite{amato1}
have found a general exact solution of the system of equations 
describing the diffusion-convection of accelerated particles, and the
dynamics and thermodynamics of plasmas in the shock region, for
an arbitrary choice of the spatial and momentum dependence of the
diffusion coefficient. In the present paper we use the formalism 
proposed by \cite{amato1} and combine it with calculations of the
perturbations created through streaming instability, so that the
diffusion coefficient, as a function of spatial location and momentum,
is determined from the spectrum and spatial distribution of the 
accelerated particles. This provides the first combined description
of the process of particle acceleration at collisionless shocks 
in the presence of particle reaction and wave generation. In this
approach, the spectrum of accelerated particles, their distribution 
in the upstream plasma and the diffusion coefficient are outputs
of the problem.

The paper is organized as follows: in Sec.~\ref{sec:solution} we
summarize the findings of \cite{amato1}. In Sec.~\ref{sec:diff}
we illustrate our treatment of the streaming instability and 
determine a relation between the power spectrum of magnetic fluctuations
and the diffusion coefficient upstream. In Sec.~\ref{sec:full} we
describe the results of our calculations. In Sec.~\ref{sec:heating} we
shortly discuss how the results presented in the previous section change
when the effects of turbulent heating are taken into account. 
We conclude in Sec.~\ref{sec:concl}.

\section{Calculations of the spectrum for arbitrary diffusion coefficient}
\label{sec:solution}

In this section we briefly summarize the mathematical procedure proposed
by \cite{amato1} to calculate the spectrum and spatial distribution 
of particles accelerated at astrophysical shocks, and their dynamical 
reaction on the shock structure, for an arbitrary diffusion coefficient
$D(x,p)$. The reader is referred to the paper by \cite{amato1} for 
more details.

The equation for the conservation of the momentum between upstream 
infinity and a point $x$ in the upstream region can be written as:
\begin{equation}
\xi_c (x) = 1 + \frac{1}{\gamma_g M_0^2} - U(x) - \frac{1}{\gamma_g M_0^2}
U(x)^{-\gamma_g},
\label{eq:normalized1}
\end{equation}
where $\xi_c (x) = P_{CR}(x)/\rho_0 u_0^2$ and $U(x)=u(x)/u_0$ and we
used conservation of mass $\rho_0 u_0 = \rho(x) u(x)$ (here $\rho_0$
and $u_0$ refer to the density and plasma velocity at upstream infinity, 
while $\rho(x)$ and $u(x)$ are the density and velocity at the location 
$x$ upstream. $M_0$ is the sonic Mach number at upstream infinity). 

The pressure in the form of accelerated particles is defined as
\begin{equation}
P_{CR}(x) = \frac{1}{3} \int_{p_{inj}}^{p_{max}} dp\ 4 \pi p^3 v(p) f(x,p),
\end{equation}
and $f(x,p)$ is the distribution function of accelerated particles. 
Here $p_{inj}$ and $p_{max}$ are the injection and maximum momentum. The 
function $f$ vanishes at upstream infinity, which implies that there 
are no cosmic rays infinitely distant from the shock in the upstream 
region \footnote{This assumption implies that we are not considering
any reacceleration of pre-existing seed particles.}. The distribution 
function satisfies the following transport equation in the reference 
frame of the shock:
\begin{equation}
\frac{\partial}{\partial x}
\left[ D(x,p)  \frac{\partial}{\partial x} f(x,p) \right] - 
u  \frac{\partial f (x,p)}{\partial x} + 
\frac{1}{3} \left(\frac{d u}{d x}\right)
~p~\frac{\partial f(x,p)}{\partial p} + Q(x,p) = 0.
\label{eq:trans}
\end{equation}
\footnote{Since we will be using this equation in Sect. \ref{sec:full} for
  the case in which diffusion is due to a strongly amplified turbulent
  magnetic field, a few comments are in order: rigorously, this
  equation describes the isotropic part of the distribution function, 
  and as long as the quasi-linear theory holds, the anisotropic part 
  is expected to represent a small perturbation. It is not clear how 
  the equation would generalize to the strongly non-linear case,
  though it may be reasonable to assume that the anisotropy remains
  rather small as long as the Alfven speed in the perturbed field is
  negligible compared with the fluid speed.} 

The $x$ axis is oriented from upstream infinity ($x=-\infty$) to
downstream infinity ($x=+\infty$), with the shock located at $x=0$.
The injection is introduced here through the function $Q(x,p)$. The
diffusion properties are described by the arbitrary function $D(x,p)$,
depending on both momentum and space \footnote{In writing Eq.~\ref{eq:trans}
in this form we are neglecting the velocity of the scattering centers
$u_w$ with respect to the fluid velocity upstream. This is always a good
approximation for the cases considered in this paper.}.

\cite{amato1} showed that an excellent approximation to the solution
$f(x,p)$ has the form 
\begin{equation}
f(x,p) = f_0(p) \exp\left[-\frac{q(p)}{3}\int_x^0 dx' \frac{u(x')}{D(x',p)}
\right],
\label{eq:solution}
\end{equation}
where $f_0(p)=f(x=0,p)$ is the cosmic rays' distribution function at the shock 
and $q(p)=-\frac{d\ln f_0(p)}{d \ln p}$ is its local slope in momentum
space.

The function $f_0(p)$ can be written in a very general way as found by
\cite{blasi1}:
\begin{equation}
f_0 (p) = \left(\frac{3 R_{tot}}{R_{tot} U_p(p) - 1}\right) 
\frac{\eta n_0}{4\pi p_{inj}^3} 
\exp \left\{-\int_{p_{inj}}^p 
\frac{dp'}{p'} \frac{3R_{tot}U_p(p')}{R_{tot} U_p(p') - 1}\right\}.
\label{eq:inje}
\end{equation}
Here we introduced the function $U_p(p)=u_p/u_0$, with
\begin{equation}
u_p = u_1 - \frac{1}{f_0(p)} 
\int_{-\infty}^0 dx (du/dx)f(x,p)\ ,
\label{eq:up}
\end{equation}
where $u_1$ is the fluid velocity immediately upstream (at $x=0^-$).
We used $Q(x,p) = \frac{\eta n_{gas,1} u_1}{4\pi p_{inj}^2} 
\delta(p-p_{inj})\delta(x)$, with $n_{gas,1}=n_0 R_{tot}/R_{sub}$ the 
gas density immediately upstream ($x=0^-$) and $\eta$ the fraction of 
the particles crossing the shock which are going to take part in the 
acceleration process. In the expressions above we also introduced the
two quantities $R_{sub}=u_1/u_2$ (compression factor at the subshock)
and $R_{tot}=u_0/u_2$ (total compression factor). If the heating of the
upstream plasma takes place only due to adiabatic compression, the two 
compression factors are related through the following expression 
(\cite{blasi1}):
\begin{equation}
R_{tot} = M_0^{\frac{2}{\gamma_g+1}} \left[ 
\frac{(\gamma_g+1)R_{sub}^{\gamma_g} - (\gamma_g-1)R_{sub}^{\gamma_g+1}}{2}
\right]^{\frac{1}{\gamma_g+1}},
\label{eq:Rsub_Rtot}
\end{equation}
where $M_0$ is the Mach number of the fluid at upstream infinity and 
$\gamma_g$ is the ratio of specific heats for the fluid. The parameter
$\eta$ in Eq.~\ref{eq:inje} contains the very important information 
about the injection of particles from the thermal bath. We adopt here the
recipe proposed by \cite{vannoni} that allows us to 
relate $\eta$ to the compression factor at the subshock as:
\begin{equation}
\eta = \frac{4}{3\pi^{1/2}} (R_{sub}-1) \xi^3 e^{-\xi^2}.
\end{equation}
Here $\xi$ is a parameter that identifies the injection 
momentum as a multiple of the momentum of the thermal particles in
the downstream section ($p_{inj}=\xi p_{th,2}$). The latter is an
output of the non linear calculation, since we solve exactly the modified
Rankine-Hugoniot relations together with the cosmic rays' transport 
equation. For the numerical calculations 
that follow we always use $\xi=3.5$, that corresponds to a fraction 
of order $10^{-4}$ of the particles crossing the shock to be injected 
in the accelerator. 

In terms of the distribution function (Eq.~\ref{eq:solution}), we can 
also write the normalized pressure in accelerated particles as:
\begin{equation}
\xi_c (x) = \frac{4\pi}{3\rho_0 u_0^2} \int_{p_{inj}}^{p_{max}} dp\ p^3
v(p) f_0(p) \exp\left[ -\int_x^0 dx' \frac{U(x')}{x_p(x',p)}
\right],
\label{eq:normalized2}
\end{equation}
where for simplicity we introduced $x_p(x,p)=\frac{3D(p,x)}{q(p) u_0}$.

By differentiating Eq.~\ref{eq:normalized2} with respect to $x$ we obtain
\begin{equation}
\frac{d\xi_c}{dx} = \lambda(x) \xi_c(x) U(x),
\label{eq:differ}
\end{equation}
where
\begin{equation}
\lambda(x)=<1/x_p>_{\xi_c}=
\frac{\int_{p_{inj}}^{p_{max}} dp~p^3 \frac{1}{x_p(x,p)} v(p) f_0(p) 
\exp\left[ -\int_x^0 dx' \frac{U(x')}{x_p(x',p)}\right]}
{\int_{p_{inj}}^{p_{max}} 
dp~p^3 v(p) f_0(p) \exp\left[ -\int_x^0 dx' \frac{U(x')}{x_p(x',p)}\right]},
\label{eq:lambda}
\end{equation}
and $U(x)$ is expressed as a function of $\xi_c(x)$ through 
Eq.~\ref{eq:normalized1}. 

Finally, after integration by parts of Eq.~\ref{eq:up}, one is
able to express $U_p(p)$ in terms of an integration involving $U(x)$ alone:
\begin{equation}
U_p(p) = \int_{-\infty}^0 dx\ U(x)^2 \frac{1}{x_p(x,p)}
\exp\left[ -\int_x^0 dx' \frac{U(x')}{x_p(x',p)}\right]\ ,
\label{eq:up2}
\end{equation}
which allows one to easily calculate $f_0(p)$ through Eq.~\ref{eq:inje}. 

Eqs.~\ref{eq:normalized1} and \ref{eq:differ} can be solved by iteration 
in the following way: for a fixed value of the 
compression factor at the subshock, $R_{sub}$, the value of the
dimensionless velocity at the shock is calculated as $U(0)=R_{sub}/
R_{tot}$. The corresponding pressure in the form of accelerated
particles is given by Eq.~\ref{eq:normalized1} as 
$\xi_{c}(0) = 1 + \frac{1}{\gamma_g M_0^2} -\frac{R_{sub}}{R_{tot}}
- \frac{1}{\gamma_g M_0^2} \left(\frac{R_{sub}}{R_{tot}}\right)^{-\gamma_g}$.
This is used as a boundary condition for Eq.~\ref{eq:differ}, where
the functions $U(x)$ and $\lambda(x)$ (and therefore $f_0(p)$) on the 
right hand side at the $k^{th}$ step of iteration are taken as the
functions at the step $(k-1)$. In this way the solution of 
Eq.~\ref{eq:normalized1} at the step $k$ is simply
\begin{equation}
\xi_c^{(k)}(x) = \xi_c(0) \exp\left[-\int_x^0 d x' 
\lambda^{(k-1)}(x') U^{(k-1)}(x')\right],
\end{equation}
with the correct limits when $x\to 0$ and $x\to -\infty$. At each 
step of iteration the functions $U(x)$, $f_0(p)$, $\lambda(x)$ are
recalculated (through Eq.~\ref{eq:normalized1}, 
Eqs.~\ref{eq:up2} and \ref{eq:inje}, and Eq.~\ref{eq:lambda}, respectively), 
until convergence is reached. The solution of this set of equations, 
however, is also a solution of our physical problem only if the pressure
in the form of accelerated particles as given by Eq.~\ref{eq:normalized1} 
coincides with that calculated by using the final $f_0(p)$ in 
Eq.~\ref{eq:normalized2}. This occurs only for one specific
value of $R_{sub}$, which fully determines the solution of
our problem for an arbitrary diffusion coefficient as a function 
of location and momentum.

\section{Self-generated turbulence and particle diffusion}
\label{sec:diff}

The streaming of cosmic rays at super-Alfv\'enic speed induces a
streaming instability, which has been discussed in previous literature
(e.g. \cite{bell78a}).

Let us define $\mathcal{F}(x,k)$ as the energy density per unit 
logarithmic band width of waves with wave-number $k$. Neglecting the 
damping, and assuming a steady state, the following relation holds 
(see e.g. Lagage \& Cesarsky 1983):
\begin{equation}
u \frac{\partial \mathcal{F}(x,k)}{\partial x} =\sigma(x,k) \mathcal{F}(x,k)\ ,
\label{eq:waven}
\end{equation}
where $u=u(x)$ is the fluid velocity upstream of the shock and 
$\sigma$ is the growth rate of waves with given wavenumber $k$, which
can be related to the distribution function of the resonant cosmic
rays, $f(x,p(k))$, through: 
\begin{equation}
\sigma=\frac{4\ \pi}{3}\ \frac{v_A}{U_M \mathcal{F}}\ \left[p^4 v
\frac{\partial f(x,p)}{\partial x}\right]_{p=\bar p(k)}\ .
\label{eq:sigma}
\end{equation}
In Eq.~\ref{eq:sigma}, $v$ and $p$ are the particle 
velocity and momentum respectively, and the latter is related to the 
wave number $k$ through the resonance condition $\bar p(k)=e B/k m c$, 
$U_M$ is the energy density of the background magnetic field 
$B_0$ ($U_M=B_0^2/8 \pi$), while $v_A$ is the local Alfv\'en velocity. 

All these expressions have been actually obtained for shocks that are
not modified by the dynamical reaction of cosmic rays. In principle,
the Fourier analysis used to obtain the previous expressions and in
fact used to reach the conclusion that there are unstable modes, is
not formally applicable, since all these calculations assume that the
background quantities (the fluid velocity $u$ in particular) are
spatially constant. However, provided that $1/k$ remains much
smaller than the spatial extension of the precursor, the conclusions
are, in first approximation, still applicable. Clearly this condition
is broken by definition at the maximum momentum $p_{max}$ at least in
those cases in which this is determined by the finite size of the
accelerator rather than by energy losses. Special care should be taken
of the fact that all quantities involved in the equations above depend
on the location in the presursor. 

It follows that for a cosmic ray modified shock, $v_A$ is not spatially
constant since both the upstream plasma density, $\rho$, and, in general, 
the background magnetic field, $B_0$, are space dependent. However all
previous calculations apply to the case of a parallel shock, for which
the strength of the background magnetic field $B_0$ can be taken to be
constant, since there is no adiabatic compression of the magnetic
field lines.  

Using the equation for conservation of mass $\rho(x)=\rho_0/U(x)$, 
we can therefore write the local Alfv\'en velocity as:
\begin{equation}
v_A(x)=\frac {B_0}{\sqrt{4 \pi \rho_0}} U(x)^{1/2}\ .
\label{eq:va}
\end{equation}

An additional warning should be issued in that Eq. \ref{eq:waven} 
neglects the adiabatic compression of waves in the shock precursor:
this reflects in the absence of terms proportional to the gradient of
the velocity field. Unfortunately, to our knowledge, discussions of
this problem in the literature are limited to integrated
quantities (e.g. total energy density and pressure of the waves) while a
description of the behaviour of the modes with different wave-numbers
is more complex. In fact, in principle even the concept of modes with
given $k$ becomes ill defined in a background which has spatial
gradients of the quantities to be perturbed.

Once $\mathcal{F}(x,k)$ is known, the diffusion coefficient is known in
turn (\cite{bell78a}):
\begin{equation}
D(x,p)=\frac{4}{\pi}\ \frac{r_L\ v}{3\ \mathcal{F}}\ .
\label{eq:diff}
\end{equation}
From the latter equation, where $r_L$ stands for the Larmor radius of 
particles of momentum $p$, it is clear that the diffusion coefficient tends 
to Bohm's expression for $\mathcal{F} \rightarrow 1$. On the other hand, it
is also straightforward to check what the expected saturation level for 
the overall energy density of the perturbed magnetic field is. If we 
define 
\begin{equation}
\frac{\delta B^2}{8 \pi}=\mathcal{I}=\frac{B_0^2}{8 \pi}\ 
\int \frac{dk}{k} \mathcal{F}(k)\ ,
\label{eq:Idef}
\end{equation}
from Eq.~\ref{eq:waven} and Eq.~\ref{eq:sigma} we see that:
\begin{eqnarray}
u \frac{d \mathcal{I}}{dx}&=&\frac{B_0^2}{8 \pi} \int \frac{dk}{k}\ 
\sigma\ \mathcal{F}(k,x)=\nonumber \\
&=&\frac{4\ \pi}{3}\ v_A \frac{d}{dx} \int dp\ v(p)\ p^3\ f(x,p)=\\
&=&v_A \frac{dP_{CR}}{dx}\ . \nonumber
\label{eq:perten}
\end{eqnarray}
Integration of the latter equation is straightforward when non-linear 
effects on the fluid are neglected so that $u$ and $v_A$ are both 
spatially constant. One obtains $\delta B^2/8 \pi=(v_A/u) P_{CR}$, or, in
terms of amplification of the ambient magnetic field:
\begin{equation}
\left(\frac{\delta B}{B_0}\right)^2=2\ M_A\ \frac{P_{CR}}{\rho_0 u_0^2}\ ,
\label{eq:ampl}
\end{equation}
with $M_A=u_0/v_A$ the Alfv\'enic Mach number. 

It is worth stressing that for $P_{CR}/\rho_0 u_0^2\sim 1$ and $M_A\gg
1$, the predicted amplification of the magnetic field exceeds
unity. In fact, this result was initially obtained in the context of the
so-called quasi-linear theory, therefore it should be taken with
caution and checked versus numerical calculations of the non-linear
phase of amplification of the waves. It seems clear, however, that the
growth may well enter this non-linear regime and lead to turbulent
fields in the shock vicinity that exceed the pre-existing background
magnetic field. 

Let us now go back to Eqs.~[\ref{eq:waven}-\ref{eq:diff}] with the aim of
recasting the relation between the diffusion coefficient and the cosmic ray
distribution function in a more compact form. Using 
Eqs.~[\ref{eq:waven}-\ref{eq:va}], we can write:
\begin{equation}
\mathcal{F}(x,p)=\frac{8\ \pi}{3}\ \frac{v\ p^4\ \Phi(x,p)}{\rho_0\
  u_0\ v_{A0}}\ , 
\label{eq:phidef}
\end{equation}
where $v_{A0}=B_0/\sqrt{4 \pi \rho_0}$ is the Alfv\'en speed at
upstream infinity and 
\begin{equation}
\Phi(x,p)=\int_{-\infty}^x \frac{dx'}{U(x')^{1/2}}\ 
\frac{\partial f}{\partial x'}(x',p)\ .
\label{eq:phi}
\end{equation}
With this definition of $\Phi$, from Eq.~\ref{eq:diff}, we obtain:
\begin{equation}
D(x,p)=\frac{3}{2 \pi^2}\ D_{B0}\ \frac{n_0}{p^3\ \Phi(x,p)}\
\frac{v_{A0}}{c}\ \frac{u_0}{c}\ ,
\label{eq:fdiff}
\end{equation}
with $D_{B0}=m_p c^3/3eB_0$ a constant. 

It is important to notice that since the constant $D_{B0}$ is
inversely proportional to the strength of the background field $B_0$,
and the Alfven speed $v_{A0}$ is proportional to $B_0$, the diffusion
coefficient in Eq. \ref{eq:fdiff} turns out to be independent of
$B_0$. This result holds only within the context of quasi-linear
theory. Even in the context of a quasi-linear theory of the
development of magnetic perturbations, a dependence on $B_0$ could be
introduced through the quantity $\Phi$, which is affected by the laws
of conservation of momentum and energy in the precursor. However, in
the cases of interest for us we will show below that these effects are
fully negligible.

\section{Spectra of the accelerated particles and self-generated
  diffusion coefficient}
\label{sec:full}

Following the mathematical procedure outlined in Sec. \ref{sec:solution}
and in Sec. \ref{sec:diff} we are able to determine self-consistently
the spectrum of accelerated particles and the diffusion
coefficient. Within the obvious limitation of using quasi-linear
theory to calculate the diffusion coefficient for the non-linear case, 
this is the first attempt at determining the space and momentum
dependence of the diffusion coefficient together with the spectrum of
accelerated particles. While in a time-dependent approach to the
problem it would be possible to estimate the maximum energy in a
self-consistent way, here we assume for simplicity that the maximum
momentum is a given parameter. We chose to carry out the calculations
presented in the following for $p_{max}=10^5\rm m c$. 

\begin{figure}
\resizebox{\hsize}{!}{
\includegraphics{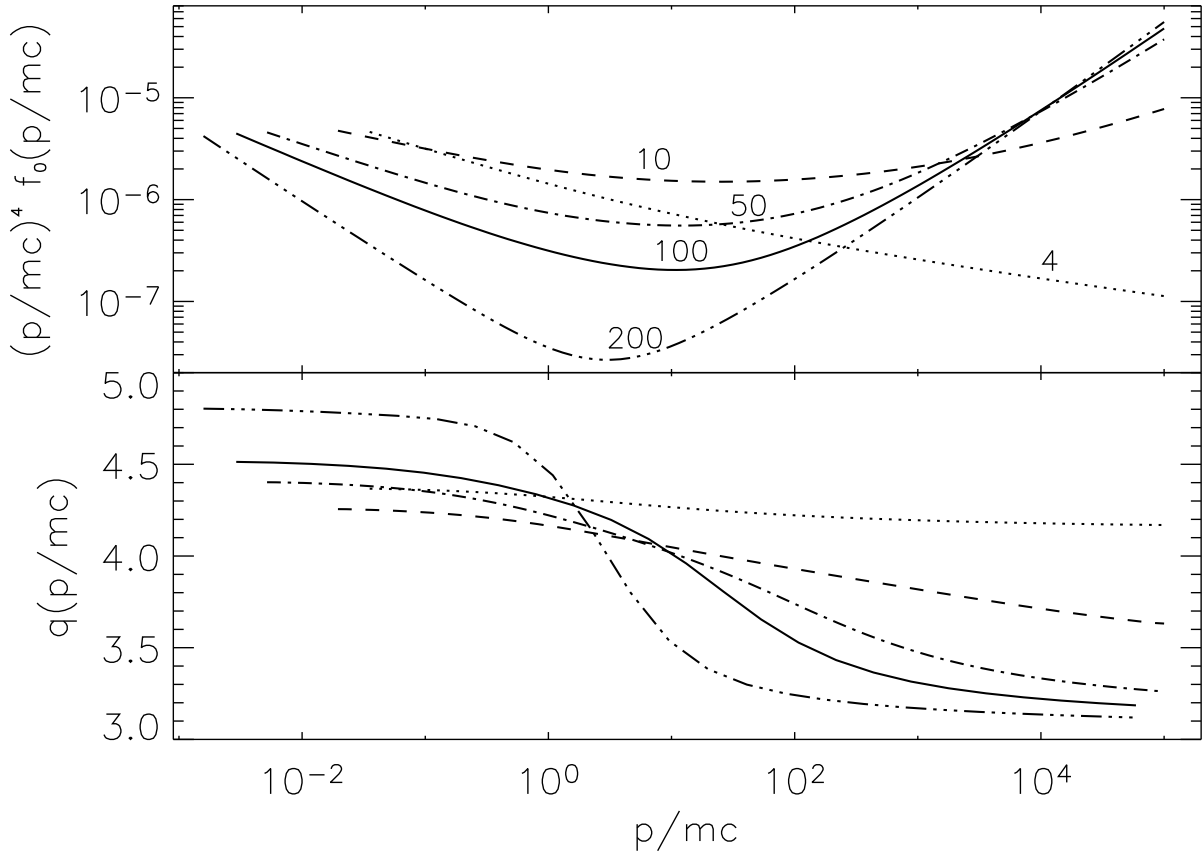}
}
\caption{Spectrum and slope at the shock location as functions of 
  energy for $p_{max}=10^5\rm m c$ and magnetic field at upstream 
  infinity $B_0=1\mu G$. The curves refer to Mach numbers at upstream 
  infinity ranging from $M_0=4$ to $M_0=200$: dotted for $M_0=4$, 
  dashed for $M_0=10$, dot-dashed for $M_0=50$, solid for $M_0=100$ 
  and dot-dot-dashed for $M_0=200$.}
\label{fig:varymn}
\end{figure}

The spectra of the accelerated particles for Mach numbers at upstream
infinity ranging from $M_0=4$ to $M_0=200$ are shown in
Fig.~\ref{fig:varymn} for a background magnetic field at upstream
infinity $B_0=1\mu G$. As stressed above the result is however
expected and actually found to be independent of the strength of 
the background magnetic field. In the bottom part of the same figure 
we plot the slope of the spectrum as a function of momentum. 

It is evident that for low Mach numbers and at given $p_{max}$ the
modification of the shock due to the reaction of the accelerated
particles is small (see for instance the case $M_0=4$). For the
strongly modified case (e.g. $M_0=200$) the asymptotic spectrum of the
accelerated particles is very flat, tending to $p^{-\alpha}$ with
$\alpha=3.1-3.2$ for $p \rightarrow p_{max}$. 
The momentum at which the spectrum becomes flatter than $p^{-4}$, the 
prediction of linear theory, depends on the level of shock
modification: it is higher ($10-20~mc$ ) for relatively low Mach
numbers (namely weaker modification) and approaches a few GeV for
high Mach numbers and large shock modification. The asymptotic
spectrum is reached at $p/mc>10^2$. These effects might be important
in the perspective of reconciling the concave shape of the
instantaneous spectra of accelerated particles with observations of
the diffuse spectrum of cosmic rays in the Galaxy. Most measurements,
mainly related to the abundance of light elements are in fact limited
to relatively low energies, where the spectra predicted in this paper 
are compatible with power laws softer than $p^{-4}$. Serious work
aimed at predicting the actual spectrum of cosmic rays escaping the
sources is urgently needed but still missing in the context of
non-linear theories of particle acceleration at shocks.  

\begin{figure}
\resizebox{\hsize}{!}{
\includegraphics{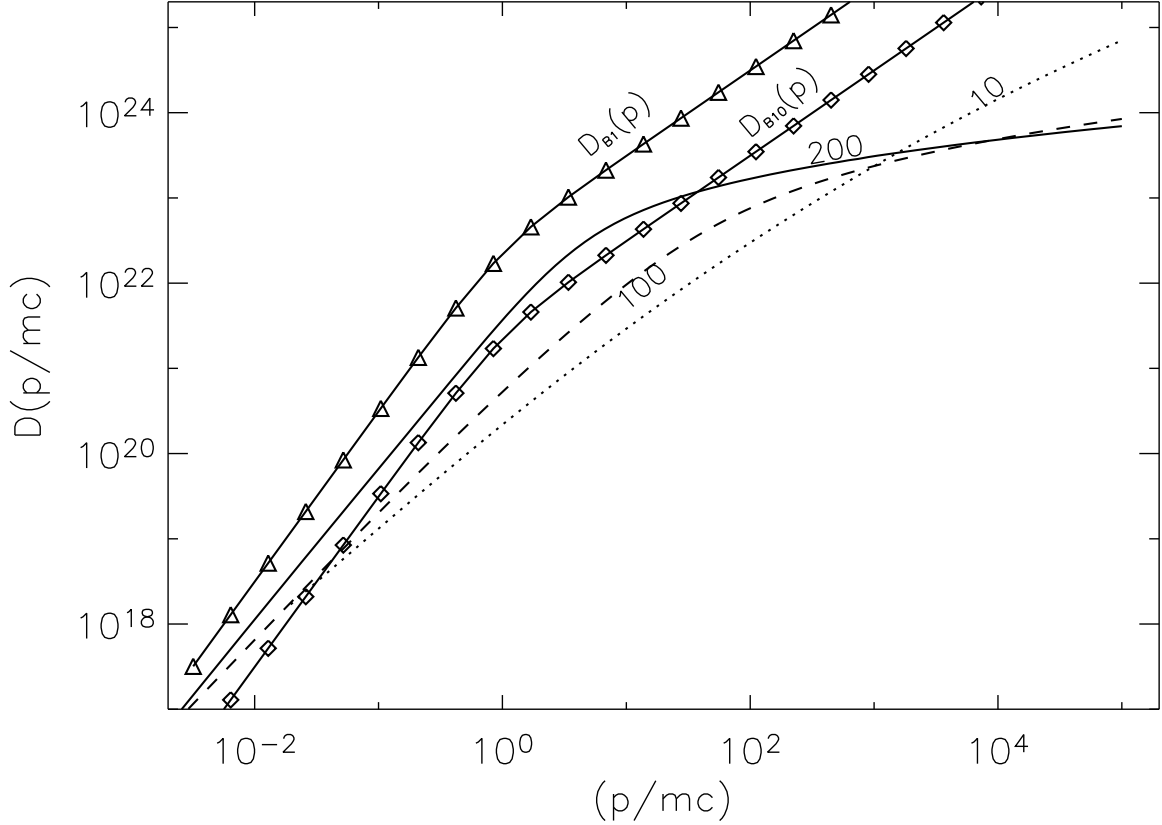}
}
\caption{The self-generated diffusion coefficient at the shock location
  $x=0^-$ as a function of the particle momentum for Mach numbers
  $M_0=10$ (dotted line), $M_0=100$ (dashed line) and $M_0=200$ (solid
  line). Also plotted is the Bohm diffusion coefficient corresponding 
  to $B_0=1 \mu G$ (solid line with triangles) and $B_0=10\mu G$
  (solid line with diamonds). The $y$-axis is in units of ${\rm cm}^2
  {\rm s}^{-1}$.} 
\label{fig:diff}
\end{figure}

The diffusion coefficient associated with the self-generated waves is
given by Eq.~\ref{eq:fdiff}. We plot this diffusion coefficient at the
shock location in Fig.~\ref{fig:diff} for Mach numbers $M_0=10$ (dashed 
lines) and $M_0=100$ (solid lines). We fix $B_0=1\mu G$, but as
stressed in the previous section, the diffusion coefficient obtained
within the quasi-linear theory of magnetic perturbations is
independent of $B_0$. For comparison, we also plot the corresponding Bohm 
diffusion coefficient $D_B(p)\propto v(p)p$ in the unperturbed
magnetic field $B_0$, for $B_0=1\mu G$ and
$B_0=10\mu G$. The comparison strikingly shows that for most momenta
of the accelerated particles the diffusion takes place at super-Bohm
rates (namely the diffusion is slower than predicted by the Bohm
coefficient in the unperturbed magnetic field, as could be expected). 
Moreover, the difference between the self-generated 
diffusion coefficient and the Bohm coefficient increases at the 
highest momenta, which might suggest that somewhat higher energies 
could be achieved if the self-generated turbulence were taken into
account. In this respect it is also important to notice that, as one
might expect, the more modified the shock, the slower the diffusion. 

It is worth keeping in mind that the diffusion coefficient in the
amplified magnetic field, as obtained through our calculations,
remains larger than the Bohm value in the same field. The latter 
is in fact considered as a sort of lower limit to the diffusion rate
(\cite{casse}) even in the case of strong turbulence. The only region in
momentum space where this condition may be violated in our
calculations is very close to the maximum momentum $p_{max}$. It is
clear however that a realistic determination of the diffusion 
coefficient cannot be achieved in the context of quasi-linear theory 
and that even numerical approaches to diffusion, such as those of 
\cite{casse} can only suggest a general trend as long as the turbulent 
structure of the magnetic field is pre-defined rather than determined 
by the diffusing particles themselves. In this sense, the limit at 
the Bohm value in the amplified field should also be taken with caution.

\begin{figure}
\resizebox{\hsize}{!}{
\includegraphics{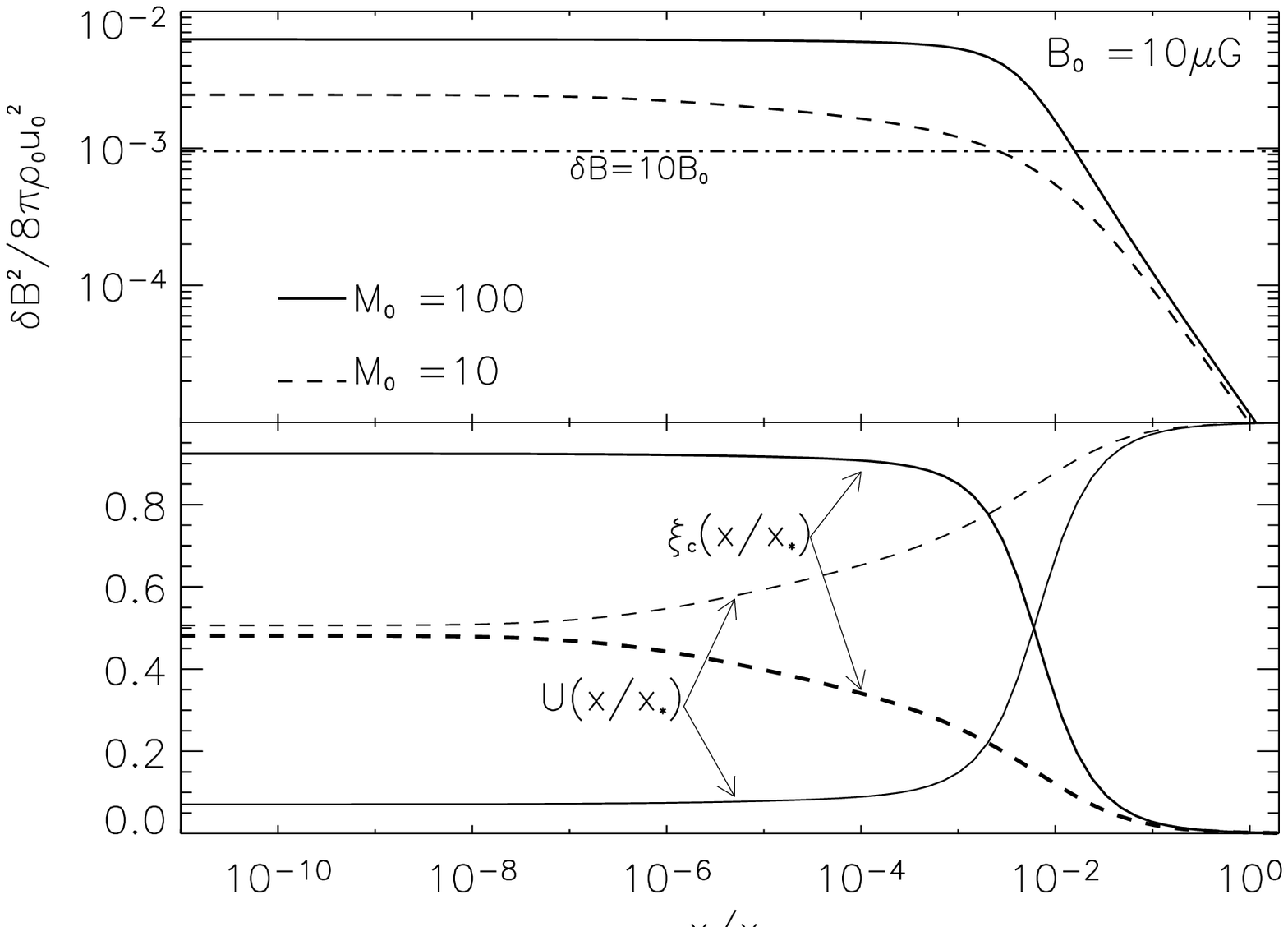}
\includegraphics{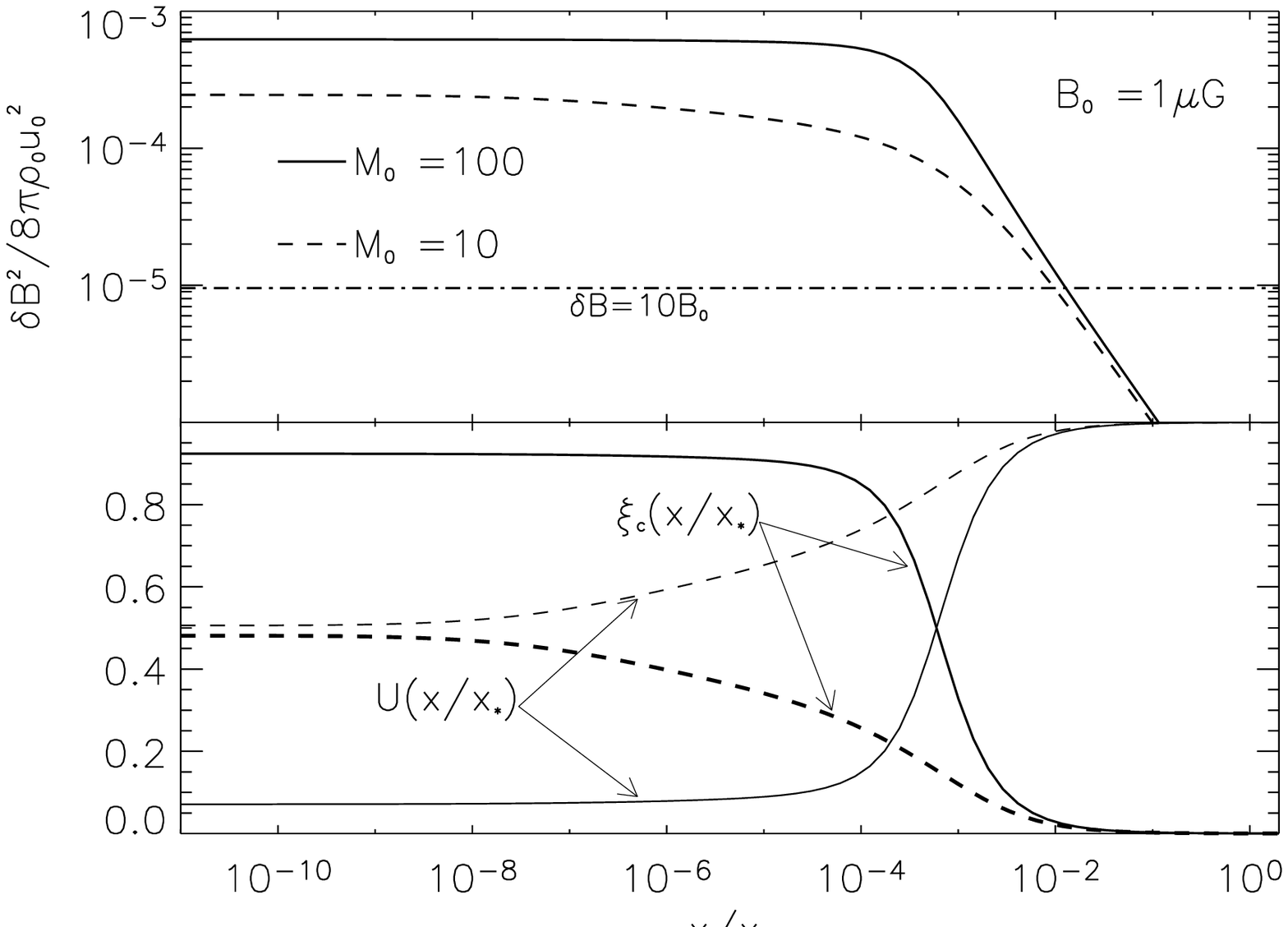}
}
\caption{Top panels: the energy density in magnetic field fluctuations 
  $\delta B^2/8\pi$ normalized to the fluid ram pressure $\rho_0 u_0^2$ 
  at upstream infinity. Bottom panels: the cosimc ray pressure normalized
  to $\rho_0 u_0^2$, $\xi_c$ (thick curves), is plotted together with the 
  normalized velocity $U$ (thin curves). All functions are plotted versus 
  spatial location, with the x-coordinate in units of 
  $x_*=-D_B(p_{\rm max})/u_0$, where $D_B(p)$ stands for the appropriate 
  Bohm diffusion coefficient. The left and right panels refer 
  to different strengths of the background magnetic field $B_0$, as specified 
  in each panel, while the different line-types correspond to different Mach 
  numbers: dashed for $M_0=10$ and solid for $M_0=100$. In the upper panels 
  we also plot for comparison a dot-dashed curve corresponding to 
  $\delta B=10 B_0$.}
\label{fig:energy}
\end{figure}

As stressed above, the fact that the diffusion coefficient is smaller 
than the Bohm coefficient in the background field is the consequence of
the fact that the fluctuations in the magnetic field become strongly 
non linear, namely $\delta B^2/B_0^2\gg 1$, at least close to the 
shock surface. In fact we find that $\delta B/B_0$ at $x=0$ is exactly
as predicted by Eq.~\ref{eq:ampl}. In these conditions it is 
important to check that the dynamical role of the turbulent magnetic
field remains small. In Fig.~\ref{fig:energy} we plot $\delta
B^2/8\pi$ normalized to $\rho_0 u_0^2$ (top panels) and the cosmic ray
normalized pressure $\xi_c(x)$  and velocity $U(x)$ (bottom panels).
The curves refer to Mach number $M_0=10$ (dashed lines) and $M_0=100$ 
(solid lines). The plots on the left (right) are obtained for 
$B_0=10\mu G$ ($B_0=1\mu G$). The $x$-coordinate is in units of 
$x_*=-D_B(p_{\rm max})/u_0$, where $D_B(p)$ stands for the Bohm 
diffusion coefficient appropriate to the considered value of $B_0$.  

The highest values of $\delta B^2/8\pi\rho_0 u_0^2$, reached close to the
shock front, are of the order of $10^{-2}-10^{-3}$, confirming that 
even in the extreme non linear cases the dynamical effect of the 
magnetic field remains unimporant. This result serves as a
justification {\it a posteriori} that we could neglect the pressure of
the waves and their energy flux in the equations of conservation of
momentum and energy respectively. This result is very specific of the
resonant channel of production of Alfven waves, and is very likely not
correct in the case of non-resonant scenarios, such as the one
proposed by \cite{bell04}, where larger amplifications of the magnetic
field could be achieved. 

\begin{figure}
\resizebox{\hsize}{!}{
\includegraphics{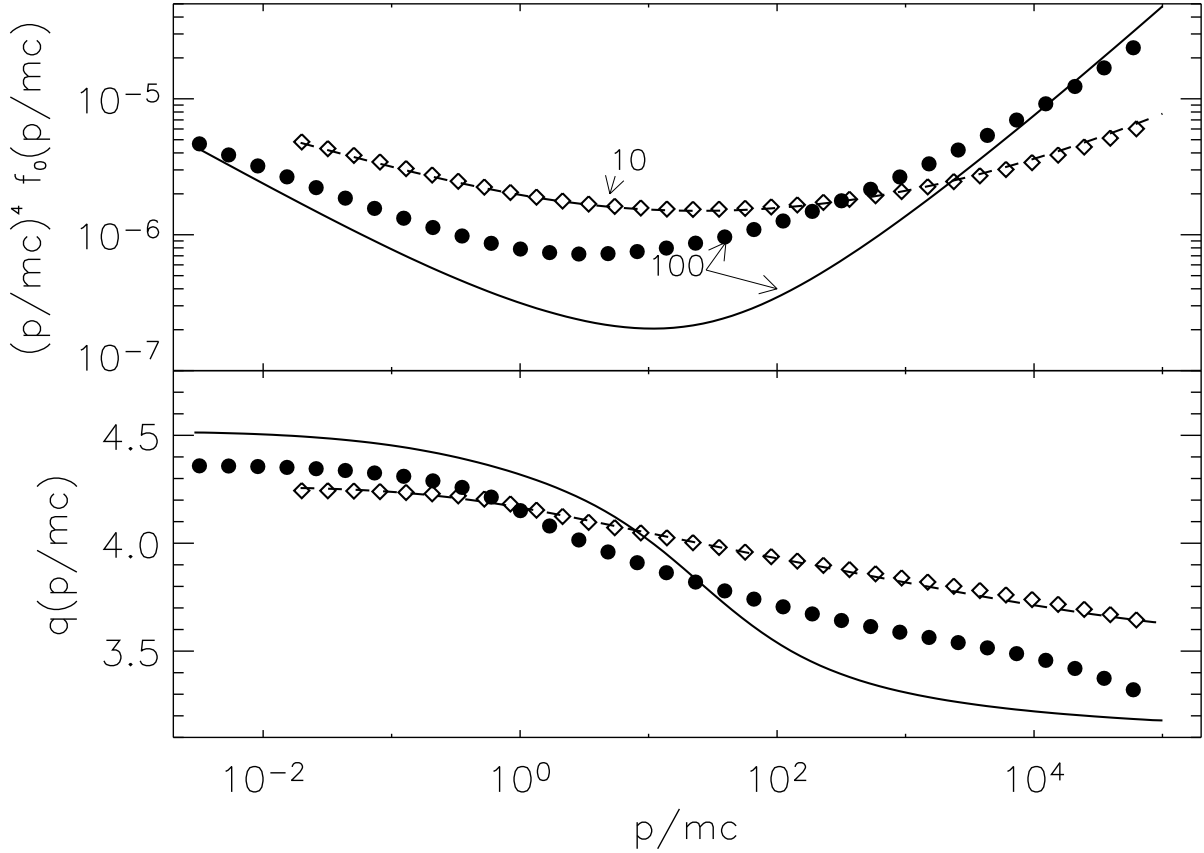}
}
\caption{Effects of using the self-generated diffusion coefficient 
  on the particles' distribution functions. In the top and bottom 
  panels we plot the particles' spectrum and spectrum slope
  respectively. The continuous
  curves are obtained for self-generated $D(x,p)$: dashed curves are for 
  $M_0=10$ and solid for $M_0=100$. The symbols represent the results 
  obtained adopting the Bohm diffusion coefficient: diamonds are for 
  $M_0=10$ and filled circles are for $M_0=100$.}
\label{fig:modspec}
\end{figure}

The shape of the spectra of accelerated particles is affected in a
sizeable way by the adoption of the self-generated diffusion
coefficient: Fig.~\ref{fig:modspec} illustrates this point. 
The continuous lines are for self-generated diffusion (dashed for
$M_0=10$ and solid for $M_0=100$) while the symbols are for Bohm
diffusion (diamonds for $M_0=10$, almost perfectly superposed
on the dashed curve, and filled circles for $M_0=100$). While in the 
weakly modified cases the spectrum is basically independent of the 
assumed diffusion coefficient, in the fully non linear solution 
self-generated diffusion leads to steeper spectra at low momenta 
and harder spectra at high momenta. 

\section{The role of turbulent heating}
\label{sec:heating}

A major uncertainty in all types of calculations of the non linear
particle acceleration at shock fronts is the effect of {\it turbulent
heating}. This generic expression is used to refer to any process that
may determine non-adiabatic gas heating. The two best known examples
of this type of processes are Alfv\'en heating (\cite{mv82}) and 
acoustic instability (\cite{drury}). Both effects are however very hard
to implement in a quantitative calculation: in the case of Alfv\'en
heating, the mechanism was originally introduced as a way to avoid the
turbulent magnetic field to grow to non linear levels, while it is
usually used even in those cases in which $\delta B/B_0\gg 1$. 

Acoustic instability develops in the pressure gradient induced by
cosmic rays in the precursor and results in the development of a train
of shock waves that heat the background gas (\cite{drury}). The analysis
of the instability is carried out in the linear regime, therefore it
is not easy to describe quantitatively the heating effect.

In both cases the net effect is the non-adiabatic heating of the gas
in the precursor, which results in the weakening of the precursor
itself and in the reduction of the acceleration efficiency compared
with the case in which the turbulent heating is not taken into
account.  

In order to illustrate this effect, we adopt a phenomenological
approach, similar to that of \cite{simple}. We stress 
that this approach, developed for the case of weak turbulence, is
inadequate in principle for the case of interest here, where the
turbulence can, in principle, become strong. We adopt it here only 
for illustration of the main physical effects. 

The approach consists in redefining the equation of state of the gas
taking into account the heating induced by the accelerated particles
(\cite{mv82}):

\begin{equation}
\frac{\partial}{\partial x} \left( P_g(x)\ \rho(x)^{-\gamma_g} \right)
= (\gamma_g-1) \frac{v_H(x)}{u(x)}\ \frac{\partial P_{CR}}{\partial x}
\rho(x)^{-\gamma_g}\ ,
\label{eq:heat}
\end{equation}
where $u/v_H=M_H$ is the local Mach number of the turbulence relevant for 
the heating (for instance $v_H=v_A$ for Alfv\'enic heating). 
After defining $\tau(x)=P_g(x)/\rho_0 u_0^2$, we replace 
Eq.~\ref{eq:normalized1} in the set of equations presented in 
Sec.~\ref{sec:solution} with the two following equations:
\begin{equation}
\xi_c(x)+U(x)+\tau(x)=1+\frac{1}{\gamma_g M_0^2}\ ,
\label{eq:newmom}
\end{equation}
\begin{equation}
\tau(x)=\frac{U(x)^{-\gamma_g}}{\gamma_g M_0^2} \left\{1+\gamma_g
(\gamma_g-1)
\frac{M_0^2}{M_{H0}}\left[\frac{1-U(x)^{\gamma_g+s}}{\gamma_g+s}+
\mathcal{I}_\tau(x)\right]\right\}\,
\label{eq:tau}
\end{equation}

where the latter is obtained by rewriting Eq.~\ref{eq:heat} in terms of 
the normalized pressures and integrating between upstream infinity and 
a generic location $x$ in the upstream medium, after expressing
$\xi_c(x)$ in terms of $\tau(x)$ and $U(x)$ through Eq.~\ref{eq:newmom}. 
We have assumed a spatial dependence of the turbulence 
characteristic velocity $v_H$ in the form of $v_H(x)=v_{H0}\ U(x)^s$ 
(with $s=1/2$ in the case of Alfv\'en heating, from Eq.~\ref{eq:va}), 
and used as a boundary condition $\tau(-\infty)=1/(\gamma_g M_0^2)$. 
The term $\mathcal{I}_\tau(x)$ appearing in Eq.~\ref{eq:tau}, finally, 
is defined as:
\begin{equation}
\mathcal{I}_\tau(x)=-\int_{-\infty}^x 
U(x)^{\gamma_g-1+s}\ \frac{d\tau}{dx^\prime}\ 
dx^\prime\ .   
\label{eq:itau}
\end{equation}
The only other changes induced by the inclusion of turbulent heating
in our initial set of equations concern the relation between the compression
ratios $R_{tot}$ and $R_{sub}$ (Eq.~\ref{eq:Rsub_Rtot}) and the temperature
jump between downstream and upstream infinity, that reflects on the minimum
cosmic ray momentum $p_{inj}$.
In both cases the changes can be summarized in the appearence of a factor
$(1+F_H)$ with
\begin{equation}
F_H=\gamma_g(\gamma_g-1) \frac{M_0^2}{M_{H0}} 
\left\{\frac{1}{\gamma_g+s}
\left[1-\left(\frac{R_{sub}}{R_{tot}}\right)^{\gamma_g+s}\right]+
\mathcal{I}_\tau(0)\right\}\ .
\label{eq:fh}
\end{equation}
With this definition of $F_H$ we find:
\begin{equation}
R_{tot} = M_0^{\frac{2}{\gamma_g+1}} \left[ 
\frac{(\gamma_g+1)R_{sub}^{\gamma_g} - (\gamma_g-1)R_{sub}^{\gamma_g+1}}
{2 (1+F_H)}\right]^{\frac{1}{\gamma_g+1}},
\label{eq:Rsub_Rtot_new}
\end{equation}
and
\begin{equation}
T_2=T_0 \left(\frac{R_{tot}}{R_{sub}}\right)^{\gamma_g-1} (1+F_H) 
\frac{(\gamma_g+1)-(\gamma_g-1)R_{sub}^{-1}}
{(\gamma_g+1)-(\gamma_g-1)R_{sub}}\ .
\label{eq:T2_T0}
\end{equation}
The usual results (adiabatic heating) are recovered when 
$M_{H0}/M_0^2\rightarrow \infty$.
In spite of the apparent simplicity of these revised relations there are two 
complications arising in the solution of the system of equations. A minor
difficulty is that the equation relating $R_{sub}$ and $R_{tot}$ now
cannot be solved analytically due to the presence of the ratio 
$R_{sub}/R_{tot}$ in the definition of $F_H$. 
A more serious complication, instead, has to do with $\mathcal{I}_\tau$, 
which requires the knowledge of the complete solution of the problem. 
However not even this is too severe a problem to overcome within the 
framework of an iterative method, although sometimes it results in an 
appreciable slowing down of the calculation. In fact, it can be seen 
{\it a posteriori} that $\mathcal{I}_\tau$ is always negligible
compared to $(1-U^{\gamma_g+s})/ (\gamma_g+s)$.

\begin{figure}
\resizebox{\hsize}{!}{
\includegraphics{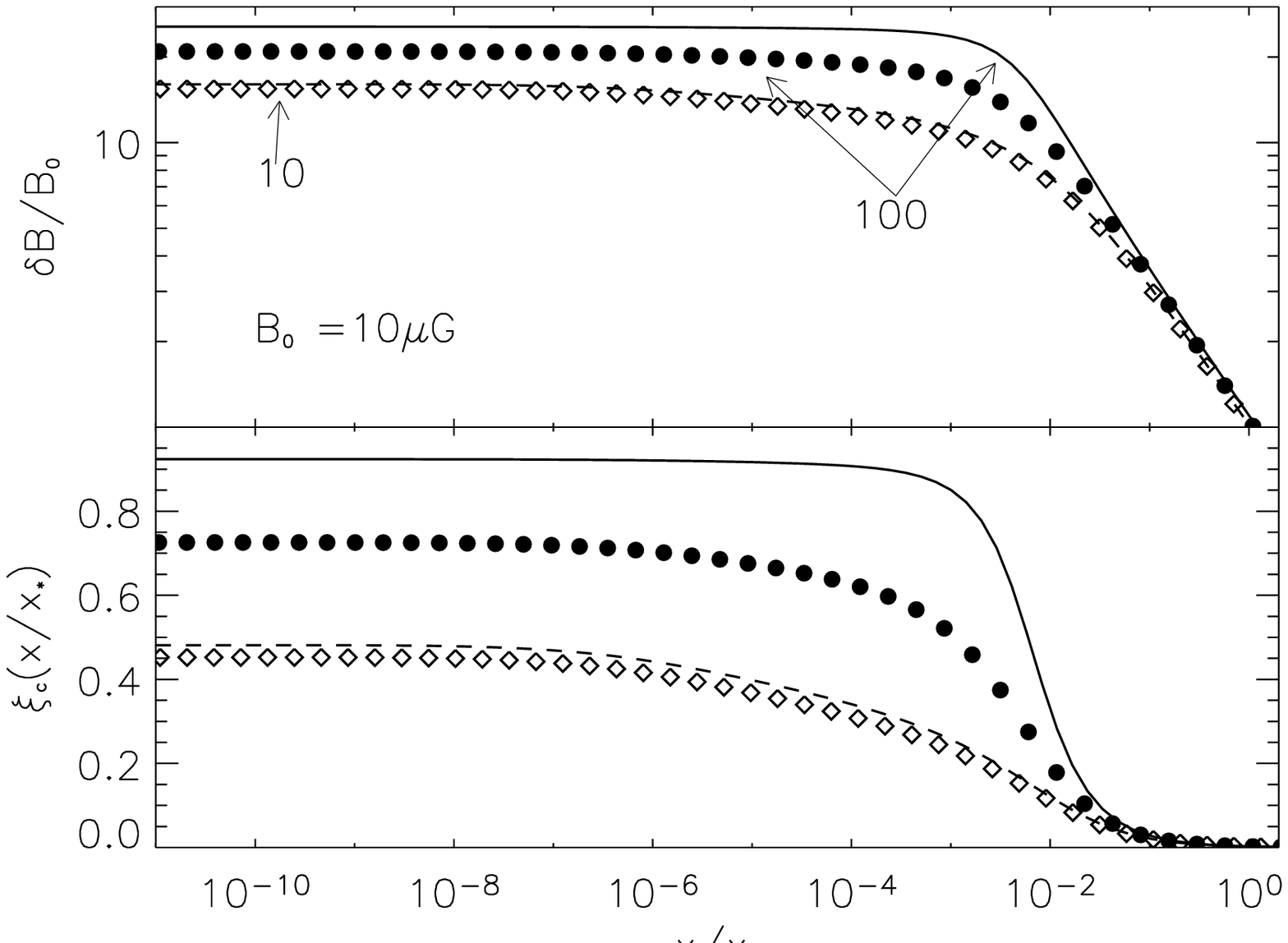}
\includegraphics{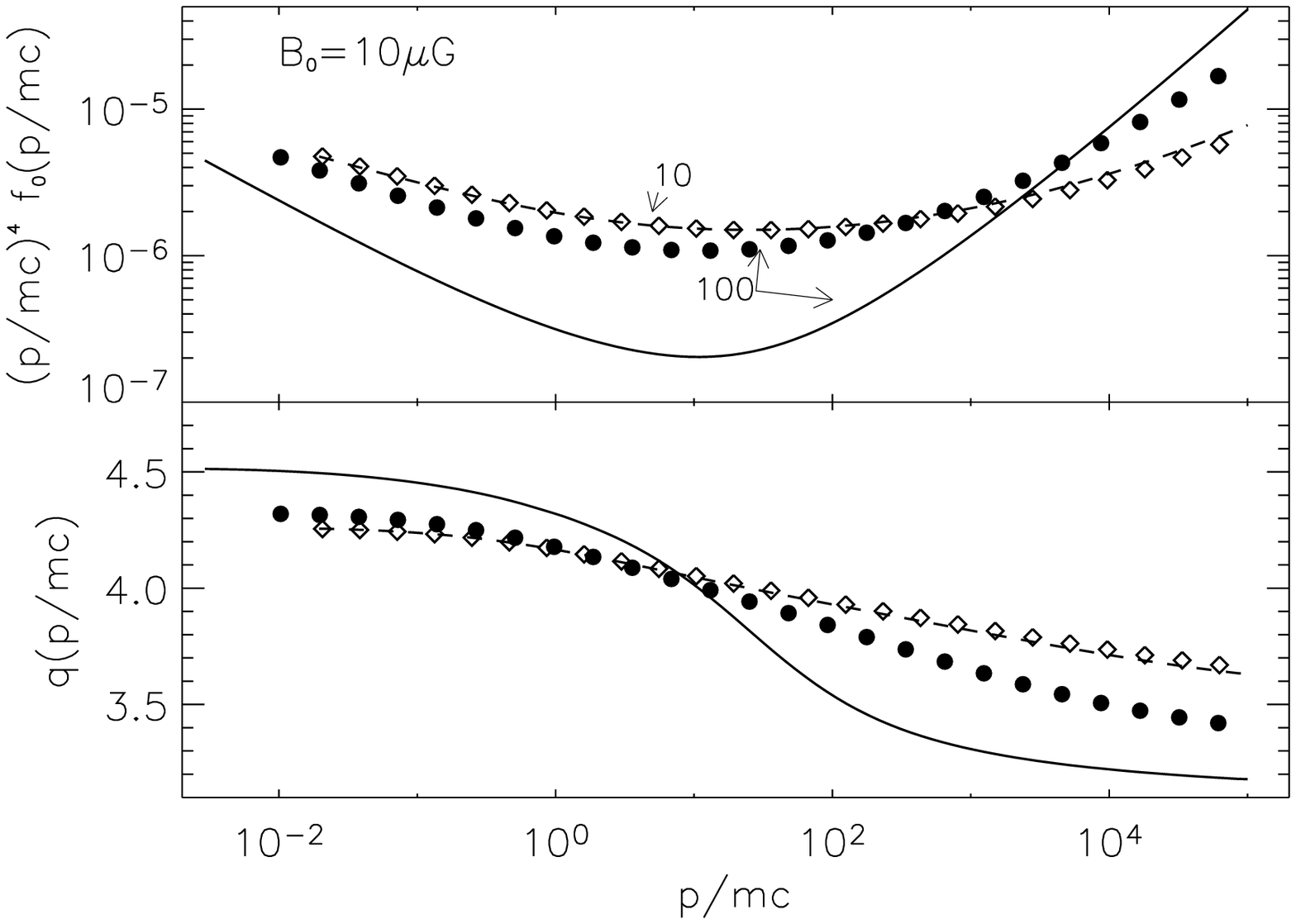}
}
\caption{The plots on the left show: in the upper panel, the ratio between 
the turbulent and background magnetic field as a function of space
for two different values of the Mach number ($10$ and $100$), 
with and without inclusion
of the turbulent heating; in the lower panel the corresponding
normalized cosmic ray pressure. The plots on the right show the particles'
spectrum and slope in the same cases. The continuous curves correspond to 
cases when the turbulent heating is not taken into account: dashed for 
$M_0=10$ and solid for $M_0=100$. The symbols correspond to cases including 
the turbulen heating: diamonds for $M_0=10$ and filled circles for $M_0=100$.}
\label{fig:heat}
\end{figure}

In order to show how our results for particle acceleration may
be affected by the inclusion of turbulent heating we carried out the 
calculations for the case of Alfv\'en heating, namely considering the 
Alfv\'en velocity as the characteristic turbulence velocity,
$v_H=v_A$, which also implies $s=1/2$ in the equations above,
according to Eq.~\ref{eq:va}. 
Results obtained with and without inclusion of Alfv\'en heating are shown
in Fig.~\ref{fig:heat}, where for the background magnetic field we have 
assumed the largest of the two values so far considered, $B_0=10 \mu G$,
with the aim of minimizing the turbulence Mach number and hence maximizing
the effects of the heating.   

In the left panel of Fig.~\ref{fig:heat} we show how the turbulent 
magnetic field strength and cosmic ray pressure are now reduced, for both 
a mildly ($M_0=10$ diamonds versus dashed line) and a strongly 
($M_0=100$, filled circles versus
solid line) modified case. In the strongly modified case we find that, in
the vicinity of the shock, the turbulent magnetic field strength is 
decreased by 10\%, while the cosmic ray pressure is decreased by 20\%. Changes
in both quantities are of order few \% in the weakly modified case.
In the right panel of the same figure we show, using the same notation for
the different curves, how the particles' spectra are affected: while changes
are negligible in the weakly modified case, for $M_0=100$ the concavity
of the spectrum is appreciably reduced, namely, the spectrum becomes
harder toward the low energy end and softer at high energies. 

It is worth stressing once more that this way of including
turbulent heating, that is used in many currect approaches to
particle acceleration in supernova remnants, is far from
self-consistent and the results should only be considered as an
indication of a trend. Sometimes, in order to attempt a slightly 
more realistic approach, one substitutes the Alfv\'en speed in the
background field $B_0$ with the corresponding quantity in the
amplified field $B_0+\delta B$. Needless to say that such an attempt,
though justified by the complete lack of any non-linear theory of
turbulent heating, is far from being realistic.

\section{Conclusions}
\label{sec:concl}

We described the mathematical theory of particle acceleration 
at non-relativistic shock fronts with dynamical reaction of
accelerated particles and self-generated scattering waves.
The diffusion coefficient itself is an output of the calculations, 
though within the limitations imposed by the usage of quasi-linear 
theory applied to the case of potentially strong magnetic field 
amplification. The scattering in the upstream plasma is generated 
through streaming instability, as discussed extensively in previous 
literature. 

We determined the spectra of accelerated particles, their spatial
distribution and the space dependence of the fluid velocity, pressure
and temperature. The diffusion coefficient and the strength of the
self-generated magnetic perturbations are also calculated, as a
function of the distance from the shock front in the precursor. 
We confirm the general finding that the spectra of accelerated
particles are concave, an effect which is particularly evident for
strongly modified shocks, namely for large Mach numbers of the moving
fluid. However, the shape of the concavity is somewhat affected by the
self-determined diffusion coefficient, as visible in
Fig. \ref{fig:modspec}.

Having in mind the comparison between the predicted spectra at the
sources and the observed cosmic ray spectrum at the Earth, it is worth 
reminding the reader that what can actually be measured is the
combination of the diffusion time, the gas density along the
trajectory (responsible for the spallation) and the injection
spectrum. In order to infer some conclusions about the spectrum at 
the source, one has to make assumptions on the diffusion coefficient 
in the interstellar medium. In alternative it would be a precious step
forward if we could measure unambiguously the spectrum of gamma rays 
generated by $\pi^0$ decays close to the source itself, an evidence
that unfortunately is still missing.

The asymptotic slope of the spectra for $p\to p_{max}$ may be as 
flat as $\sim 3.1-3.2$, but this conclusion is not strongly affected
by the fact that the diffusion coefficient is calculated self-consistently. 

The most striking new result of our calculations is the energy
dependence of the diffusion coefficient and the strength of the
amplified turbulent magnetic field. As could be expected, the 
diffusion coefficient is not Bohm-like, and the turbulent component 
of the magnetic field is amplified so efficiently that the 
diffusion coefficient is much smaller than the Bohm coefficient in the
background magnetic field.  
This is especially true at the highest momenta, which leads to
think that a full non-linear theory might predict higher values
of the maximum momentum than expected on naive grounds. Unfortunately
a full, self-consistent calculation of $p_{max}$ for a strongly
modified shock has never been carried out, the main difficulty
being in accounting for the spatial dependence of all the quantities
involved.

When compared with the Bohm diffusion coefficient as calculated
in the amplified magnetic field, our diffusion coefficient remains always
larger, with the possible exception of a narrow momentum region close
to $p_{max}$. 

While the calculation presented here is fully self-consistent in the
determination of the shock modification due to the reaction of the
accelerated particles, the part related to the amplification of the
background field suffers from all the limitations related to the usage
of quasi-linear theory for the streaming instability. This approach,
initially developed for weakly amplified magnetic fields, is widely
applied in the literature to situations that violate this
condition. Unfortunately at the present time this is the only way we
have to achieve a (at least partially) self-consistent picture of the
process of particle acceleration at cosmic ray modified shocks with
self-generated turbulence. This problem is in fact even more serious
for those approaches that predict levels of magnetic field
amplification which are much higher than those found here
(e.g. \cite{lb01,bell04}).

The high acceleration efficiencies obtained in the context of
all approaches to particle acceleration at shocks are known to be
reduced by the effect of turbulent heating. Any non-adiabatic heating
of the gas in the precursor leads to reducing the energy channelled into
non-thermal particles at the shock. This is a serious problem, because
the effect of turbulent heating depends dramatically on the type of
mechanism that is responsible for the heating: Alfv\'en heating, often
used in the literature, is only one of these mechanisms, and not
necessarily the most efficient. For instance, the instability induced 
by the propagation of acoustic waves in the precursor was shown to 
lead to the formation of weak shocks in the precursor, which in turn 
heat the upstream plasma (e.g. \cite{drury}).
 
These non-linear effects can hardly be taken into account in a
credible way. Most notably, the phenomenological expressions proposed
in the literature and used also in the present paper, have originally
been proposed as mechanisms to reduce the amount of magnetic field
amplification and remain in the context of small perturbations of the
background magnetic field. However, as shown in Fig. 5, even with the 
Alfv\'en heating taken into account, the magnetic field can be
amplified by a factor in excess of $\sim 10$ with respect to
the background field. This means that a fully non-linear theory of
the turbulent heating is required in order to make fully reliable 
predictions. 

From the phenomenological point of view, the best evidences for both 
magnetic field amplification and efficient particle acceleration come
from observations of supernova remnants (see the reviews of
\cite{hillas} and \cite{blasirev} and references therein). In fact, it
has been argued that the amount of field amplification required to
explain the thickness of the X-ray bright rims in several remnants is
of the order of $\sim 200-300\mu G$ (\cite{voelk05}). An important
role in explaining this level of amplification could be played by
different versions of the streaming instability
(\cite{lb01,bell04,bell05}), not requiring resonant interactions of
particles and waves. A full non-linear theory including these effects
will be described elsewhere (Amato and Blasi, in preparation).

\section*{Acknowledgments}
This research was funded through grant COFIN2004-2005. We wish to
acknowledge useful conversations with D. Ellison, S. Gabici and M. 
Vietri. We are also grateful to an anonymous referee for useful
comments.

\end{document}